\documentclass[useAMS,usenatbib]{mn2e}

\newcommand{\msun}{M$_{\sun}$}
\newcommand{\kms}{km s$^{-1}$}
\newcommand{\msuns}{M$_{\sun}~$}
\newcommand{\kmss}{km s$^{-1}~$}
\newcommand{\nbody}{{\it n}-body$~$}
\newcommand{\costheta}{${\rm cos}~\theta$}
\newcommand{\costhetas}{${\rm cos}~\theta~$}

\newcommand{\mnras}{MNRAS}
\newcommand{\apj}{ApJ}
\newcommand{\apjl}{ApJ}
\newcommand{\aj}{AJ}
\newcommand{\aap}{A\&A}
\newcommand{\araa}{ARA\&A}

\newcommand{\apjs}{ApJS}

\usepackage{graphicx}
\usepackage{epstopdf}
\usepackage{amssymb,amsmath}
\usepackage{subfigure}

\title[Circumbinary disc survival during scattering]{Circumbinary disc survival during binary-single scattering: towards a dynamical model of the Orion BN/KL complex}
\author[N. Moeckel and C. Goddi]{Nickolas Moeckel$^{1}$\thanks{E-mail:
moeckel@ast.cam.ac.uk} and Ciriaco Goddi$^{2}$\\
$^{1}$Institute of Astronomy, University of Cambridge, Madingley Road, Cambridge, CB3 0HA\\
$^{2}$ESO, Karl-Schwarzschild-Strasse 2, D-85748 Garching bei M{\" u}nchen
}
\begin{document}

\date{Accepted XXX. Received YYY; in original form ZZZ}

\pagerange{\pageref{firstpage}--\pageref{lastpage}} \pubyear{2009}

\maketitle

\label{firstpage}

\begin{abstract}
The Orion BN/KL complex is the nearest site of ongoing high-mass star formation. Recent proper motion observations provide convincing evidence of a recent (about 500 years ago) dynamical interaction between two massive young stellar objects in the region resulting in high velocities: the BN object and radio Source I. At the same time, Source I is surrounded by a nearly edge-on disc with radius $\sim 50$ au. These two observations taken together are puzzling: a dynamical encounter between multiple stars naturally yields the proper motions, but the survival of a disc is challenging to explain. In this paper we take the first steps to numerically explore the preferred dynamical scenario of Goddi et al., in which Source I is a binary that underwent a scattering encounter with BN, in order to determine if a pre-existing disc can survive this encounter in some form. Treating only gravitational forces, we are able to thoroughly and efficiently cover a large range of encounter parameters. We find that disc material can indeed survive a three-body scattering event if 1) the encounter is close, i.e. BN's closest approach to Source I  is comparable to Source I's semi-major axis;  and 2) the interplay of the three stars is of a short duration. Furthermore, we are able to constrain the initial conditions that can broadly produce the orientation of the present-day system's disc relative to its velocity vector. To first order we can thus confirm the plausibility of the scattering scenario of Goddi et al., and we have significantly constrained the parameters and narrowed the focus of future, more complex and expensive attempts to computationally model the complicated BN/KL region.
\end{abstract}


\section{Introduction}

The details of how massive stars form are poorly understood. 
High-mass young stellar objects (YSOs) evolve rapidly and are intrinsically rare (due to the declining initial mass function), which generally implies large distances ($\gtrsim500$~pc). Furthermore, they form deeply embedded in crowded regions where high extinction prevents observations in the optical and infrared. Lacking firm observational support, models of high-mass star formation have remained controversial, with three different scenarios active in the recent literature: (1) monolithic collapse and disc accretion \citep[e.g.][]{yorke02,mckee03}; (2) competitive accretion in protoclusters \citep[e.g.][]{bonnell01a,bonnell04}, and similar theories in which protostellar dynamics greatly affect the ongoing accretion \citep{peters10}; and (3) stellar collisions and mergers \citep{murray96,bonnell98a,bally05}, although this path is likely only viable in the most extreme clustered environments \citep{moeckel11,baumgardt11}.  

As originally formulated, the last two models are based on the evidence that most massive stars form in dense protostellar clusters \citep{lada03}, where the stellar densities may be so high that dynamical interactions (including close passages or collisions and mergers) are naturally expected and may be common in the very early stages of massive cluster evolution. In contrast, the first scenario envisions a scaled-up version of low-mass star formation, where discs and outflows are invoked to reduce the effect of radiation pressure from massive protostars and enable ongoing accretion; discs concentrate the infalling material into small solid angles and collimate the radiation field preferentially in the polar direction, where stellar photons can escape along the cavity of lower density gas excavated by outflows \citep{yorke02,krumholz09}.
In recent years some modest convergence between modelers has occurred, and the debate has advanced slightly. For instance, few today would argue that discs play no role in mediating accretion onto a massive YSO, or that radiation is unimportant. Whether the matter that ends up in a massive star is bound prior to collapse as a massive core or is subsequently gathered from its environment remains a point of vigorous discussion, however, as does the importance of (proto)stellar dynamics.
In order to constrain different theories, two different aspects should be investigated: on scales of individual YSOs, the properties of disc--outflow interfaces, in particular the structure of discs and the acceleration and collimation mechanisms of outflows; and on scales of clusters, the role of dynamical interactions among protostars in the early phases of protocluster evolution. In these respects, the Orion~BN/KL region can provide an ideal laboratory.

\begin{figure}
 \includegraphics[width=80mm]{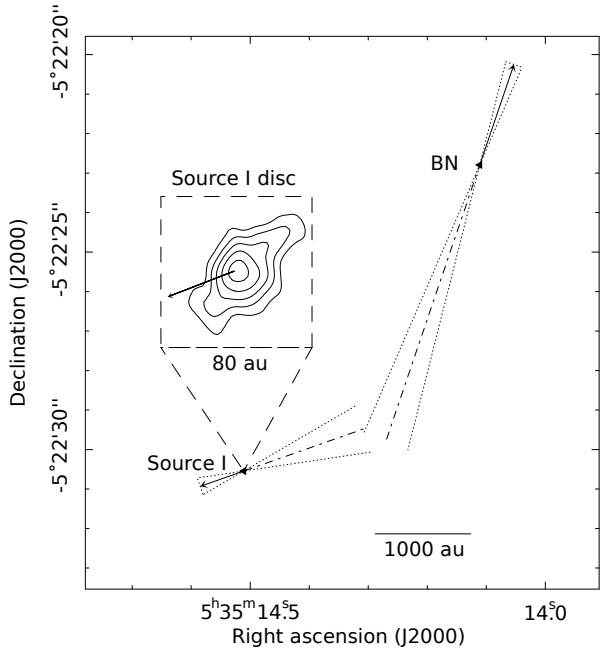}
 \caption{Proper motions of BN and Source I in the ONC rest-frame measured from multi-epoch radio continuum observations from \citet{goddi11}. The arrows indicate proper motion direction and displacement for 200 yr. The dash-dot line indicates motion backward over 560 years, assuming it is linear. The dashed lines indicate 1$\sigma$ uncertainties in position angles. The minimum (past) separation on the sky between Source I and BN was $50 \pm 100$ AU.
 The image in the inset shows the nearly edge-on compact disc around Source I as observed in 2006 with the VLA at 7mm \citep{goddi11}.
We note that another object in the region, source {\it n}, has been proposed to take part in the same dynamical interaction along with BN and Source~I \citep{gomez08}. The analysis here does not consider source {\it n},  based on the evidence reported in \citet{goddi11}.
 }
  \label{propermotions}
\end{figure}

The Orion~BN/KL complex contains the nearest region of ongoing high-mass star formation ($\sim$420~pc; \citealt{menten07}) and is a compelling target for studying how massive YSOs form and interact with their surroundings. 
A dense protostellar cluster lies within the region containing two
radio sources that are believed to be massive YSOs:  the highly embedded radio Source~I
\citep{reid07,matthews10} and the BN object \citep{becklin67}, which is the brightest source in the region in the
mid-infrared  at 12.4~$\mu$m \citep{gezari98}. 
Source I is probably the best case known of a massive protostar with ongoing disc-mediated accretion. High angular resolution monitoring of molecular masers enabled for the first time the resolution of the launch/collimation region of an outflow from a compact disc, as well as tracing the dynamical evolution of molecular gas over time at radii of 10-100 au \citep{matthews10}. The disc mass is poorly constrained, but it is estimated to lie in the range 0.002 \msun$<M_{disk}<0.2$ \msun \citep[see][]{goddi11}. Complicating matters, however, recent works report strong evidence that a dynamical interaction occurred 500~years ago between Source I and BN  \citep[e.g.][]{gomez08,goddi11}. 
The stellar interaction probably resulted in high stellar proper motions for  Source I and BN  \citep[12 and 26 \kms, respectively;][]{goddi11} and possibly produced a fast bullet outflow traced by  CO and H$_2$ emission \citep{bally11}.  In figure \ref{propermotions} we show the proper motions of Source I and BN, as well as the disc around Source I; for complete details, refer to \citet{goddi11}.

The strong evidence of a dynamical interaction and the simultaneous existence of a circumstellar disc (radius $\sim 50$~AU) around Source I are challenging to explain.
Two possibilities have been considered: an existing disc is destroyed in the encounter and then rebuilt in the following 500 years from material dispersed from the original discs or interstellar gas in the region, via Bondi-Hoyle accretion \citep{bally11}; alternatively, the disc is truncated at the radius of the close passage but survives the encounter \citep{goddi11}. The former authors speculate about the magnetoyhydrodynamic consequences of a rapidly changing potential due to chaotic stellar dynamics. The latter authors focused on the stellar dynamics of the stars alone, performing \nbody experiments to explore different dynamical scenarios, settling on a preferred model of an interaction between an existing binary and a single star.

While the full story of the complicated BN/KL region remains unclear, steps forward must involve modeling of the system.
Any modeling of the system's  dynamics must be performed from a set of initial conditions that permits a stellar interaction to: (1) result in high post-encounter velocities of the interacting stars, and (2) preserve some of the original circumstellar material despite chaotic and potentially complicated stellar dynamics. We turn to numerical experiments to attempt to constrain this set. 
In this work, we take a first-order approach to the problem, considering only the most important physical process: the gravitational interaction between the stars and its effect on the circumbinary material.  This is reasonable, as the system's mass is dominated by the stars while the gas around Source I is most likely rotationally supported.
A fully realistic approach should include, beyond gravitational dynamics, hydrodynamic, radiative, and magnetic forces. 
In that case, however, parameter space would be tremendously expanded, and likewise the computational cost of a single experiment.  
The final explanation of BN/KL will undoubtedly involve an interplay between all of these forces; we are not presenting this work as a full explanation of the region, but rather as the first step in a systematic, step-by-step approach to the problem. 

In this paper, for the first time we numerically investigate effect of a gravitational scattering encounter between stars on a circumbinary disc.
The paper is structured as follows. The details on simulations are described in \S \ref{simdetails}. Main results from the numerical work are presented in \S \ref{results}. In \S \ref{discussion}, we discuss these results in the context of existing observations of Orion~BN/KL, and in \S \ref{conclusions} we summarize our conclusions and look forward to future steps.

\begin{figure}
 \includegraphics[width=84mm]{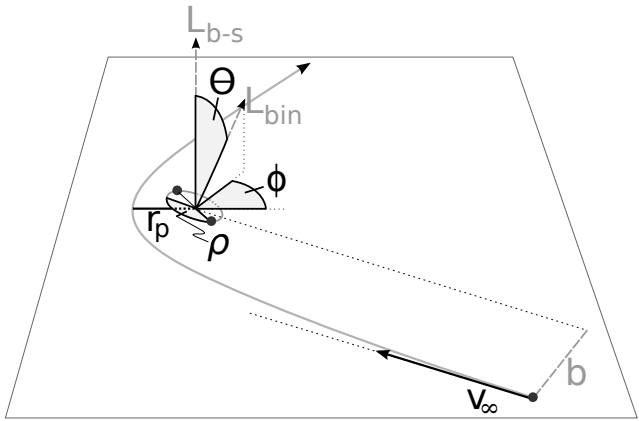}
 \caption{Schematic representation of an encounter of a single star with a binary in the binary's rest frame. The gray curved line shows the path of the hyperbolic encounter the binary and the intruder would experience in the two-body approximation where the binary is represented by its center of mass. We show the parameters we sample over (black labeled quantities), as well as some auxiliary variables that are a consequence of the initial conditions (gray labeled quantities). The periastron separation between the binary center of mass and the intruder $r_p$, their relative velocity at infinity $v_{\infty}$, and the impact parameter $b$ are related by the gravitational focusing relationship shown in equation \ref{focusing}. The angles $\theta$ and $\phi$ determine the angle between the angular momenta of the binary--single orbit, ${\bf L}_{b-s}$, and the orbit of the binary ${\bf L}_{bin}$, and $\rho$ determines the initial orbital phase of the binary.
 }
  \label{ICschematic}
\end{figure}

\section{Simulation details}
\label{simdetails}
In this work we consider an isolated \nbody system comprising of three stars, arranged as an initial binary and a single star, and a circumbinary disc, which we model with low-mass particles. 
Hydrodynamic forces in the disc are neglected, an approximation which allows us to complete many thousand scattering experiments in a reasonable timeframe. Large numbers of experiments are necessary, due to the chaos inherent to the three-body problem. It is impossible to definitively trace backwards from the current system, observed with uncertainty, to an earlier configuration; instead statistical studies must be performed to determine the likelihood of similar outcomes to what we observe.  The \nbody system is evolved using Aarseth's {\sc nbody6} code \citep{aarseth00,aarseth03}, which utilises a fourth-order Hermite integrator.  While the code includes a KS regularization scheme \citep{kustaanheimo65} for handling close encounters between bodies, because of the atypical setup of this problem (three dominant bodies and a disc of low-mass particles), we suppress this feature and the equations of motion are integrated in standard Cartesian space. Because {\sc nbody6} does not smooth the force between particles, there is in principle the possibility that undesired relaxation effects can occur in the disc; however in practice, over the short timescale of these simulations this is not a concern.

The two main elements of this problem are the interaction between the disc and the stars, and the gravitational scattering encounter between the stars. Separately, each of these elements have been extensively explored: the theory of binary-single scattering is laid out in \citet{heggie75}, and a long history of numerical work includes \citet{becker20}, \citet{saslaw74}, \citet{hills75}, \citet{valtonen79}, \citet{hut83}, and \citet{hut93}. Disruptive disc-star encounters have likewise been studied numerically in a variety of contexts \citep[e.g.][]{clarke93,heller95,hall96,boffin98,moeckel06}. The combination of the two processes is unstudied to our knowledge\footnote{Although there has been some work in a similar spirit, investigating the survival of accretion discs in black hole interactions \citep{lin77}.}, and interesting in general. Parameter space is vast, and it is fortunate that BN/KL provides a physical reference from which to take initial steps.

\begin{figure*}
 \includegraphics[width=180mm]{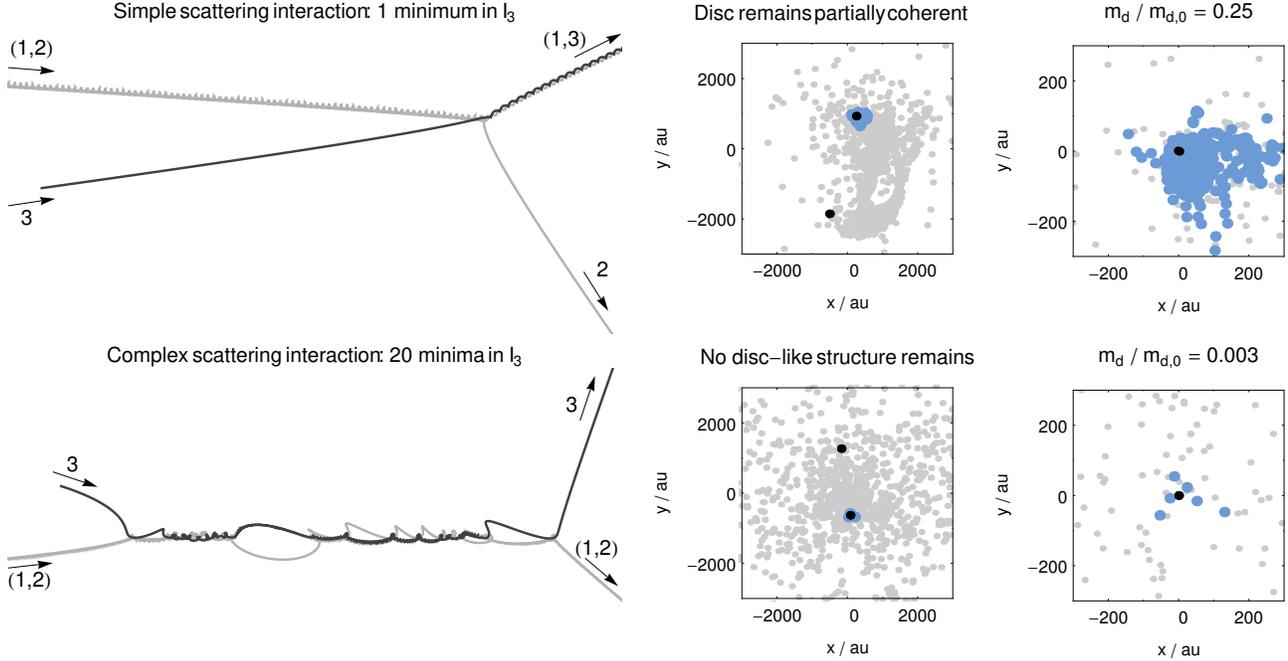}
 \caption{Examples of a successful interaction ({\it top row}) and an unsuccessful one ({\it bottom row}) from the (10,10)+10 runs.  {\it Left column}: the paths of the three stars through the interaction in a frame drifting with respect to the center of mass of the system.  The initially-single star is the darker line.  Numbers with arrows label the stars at the beginning and end of the encounter.  {\it Right columns}: the final distribution of the stars (black points) and disc particles (gray points) 500 years after the scattering event.  Disc particles considered part of the disc by the criteria discussed in the text are plotted as larger blue points. We show both a wider view in the center of mass frame, and closer view centered on the final binary.}
  \label{examples}
\end{figure*}

\subsection{Initial conditions}
We ran two sets of simulations, which we label in terms of the masses of the initial stellar configuration in \msun.  One set of 4500 simulations, (10,3)+10, has a low-mass member in the binary, while the other set of 7000 runs, (10,10)+10, has all three stars with equal masses.
Each experiment begins with a circular binary with semi-major axis $a = 10$ au surrounded by a circumbinary disc with total mass 0.1 \msun. The results of this paper are, strictly speaking, only applicable to discs that are in this non-self gravitating, dynamically unimportant regime. While the dynamics of the actual scattering interaction between the stars is probably insensitive to the disc mass, a more massive disc could potentially alter the trajectory of the stars in the immediate leadup to the stellar interactions. Note that the estimated remnant disc mass of less than 0.2 \msuns is comfortably unimportant to the dynamics of a massive binary over short timescales. The simulated disc is truncated at the inner edge at the 3:1 resonance with the binary \citep{artymowicz94} which is approximately $2.08a=20.8$ au, extends to the outer edge at $10a = 100$ au, and follows an arbitrary but plausible surface density profile $\Sigma(r) \propto r^{-1}$. Because we ignore hydrodynamic forces there is no need to resolve the disc vertically, and the disc particle are initially coplanar with the binary. The disc is composed of 2048 particles, and tests with 1024 and 4096 particles show that the results are essentially independent of disc resolution.

In order to place the (initially unbound) third star, the binary is treated as a single body at its center of mass, and a Keplerian orbit is determined for some relative velocity at infinity $v_{\infty}$ and desired periastron $r_p$.  This periastron is the unperturbed separation that occurs when the binary is treated as its center of mass; the actual orbits of the three bodies diverges from this path as they approach each other and the two-body approximation to the system breaks down, but this serves to define the initial conditions.  The third star is introduced at a separation of $110a = 1100$ au from the binary.  At this separation, the ratio of the perturber's force to the binary's force at the disc edge is less than $10^{-2}$. 

In the two-body approximation the orbit is determined by just $v_{\infty}$ and $r_p$, but there are also several angles required to identify a set of initial conditions, specifying the orientation of the binary.  These can be described by the angle between the angular momentum of the binary's orbit (${\bf L}_{bin}$) and the angular momentum of the binary--single orbit (${\bf L}_{b-s}$), given by the polar angle $\theta$ and azimuthal angle $\phi$; the orientation of the binary's Laplace-Runge-Lenz vector relative to some reference vector in the binary plane; and the mean anomaly of the binary, $M$.
The Laplace-Runge-Lenz vector is a conserved vector that describes the orientation of the binary's ellipse, and the mean anomaly is related to the position in the orbit.  Because we consider only circular binaries in this work, these are degenerate and can be replaced by the orbital phase $\rho$ of the binary, measured from the reference vector in the binary plane. For the reference vector we use ${\bf L}_{b-s} \times {\bf L}_{bin}$. In figure \ref{ICschematic} we show a schematic representation of our initial conditions

Because our primary goal is not to determine the probability of a certain scattering event \citep[as in e.g.][]{hut83} but rather to see what encounter parameters can result in disc retention, we have sampled uniformly over the periastron of the initial orbit rather than the impact parameter squared.  This means that the initial conditions we choose are not formally proportional to their probability, but it ensures that we have good coverage over potentially interesting regions of the $r_p$--$v_{\infty}$ plane that could otherwise be sampled with low probability by a standard scattering experiment setup. However, as we show below, the two sampling methods are virtually equivalent over the parameter range we cover.

\subsection{Parameter ranges}
\label{parameterranges}
Our initial conditions for each experiment are then described by five random numbers sampled with a quasi-random number generator, namely:  
\begin{align*}
r_p/{\rm au} & \in [0,60) \nonumber \\
v_{\infty}/{\rm km~s}^{-1} & \in [0,15)  \nonumber \\
 {\rm cos}~\theta & \in [-1,1) \nonumber \\
 \phi & \in [0,2\pi) \nonumber \\
 \rho & \in [0,2\pi).
\end{align*}
We have given the limits on $r_p$ and $v_{\infty}$ in physical units, but they can also be expressed in non-dimensional form relative to the natural length and velocity scales of the problem. Reasonable choices for these are the binary semi-major axis $a$, yielding $r_p/a  \in [0,6)$, and the critical velocity of the three-star system, $v_c$. This is the value of $v_{\infty}$ at which the three-body energy is zero; if $m_1$ and $m_2$ are the masses of the binary components and $m_3$ is the intruder, it is given by 
\begin{equation}
v_c = \left[ \frac{G}{a} \frac{m_1 m_2 ( m_1 + m_2 + m_3)}{m_3 ( m_1 + m_2)} \right ] ^{1/2}.
\label{vcrit}
\end{equation}
For the (10,10)+10 runs we have $v_c = 36.5$ \kmss and $v_{\infty}/{v_c}  \in [0,0.41)$, and for the (10,3)+10 runs we have $v_c = 21.7$ \kmss and $v_{\infty}/{v_c}  \in [0,0.69)$.

The upper limit for $r_p$ is determined by what might be an `interesting' encounter,  that is the relative velocity between the binary and the single is boosted during the three-body interaction. This requires a close triple approach and hardening of the binary; previous work suggests that the maximum interaction distance between the binary and the single is a few times the binary separation \citep[e.g.][]{saslaw74,valtonen74}.  Our limit of 60 au, or $6a$, safely covered the full range of potentially interesting periastra.  After an initial run of 2000 experiments, we determined that no interesting encounters were taking place beyond $r_p =30$ au for the (10,3)+10 case.  After this an additional 1500 experiments were performed, drawn from the same quasi-random sampling series but carrying out only those runs with $r_p \le 30$ au.  The (10,10)+10 case had a few interesting encounters at larger periastra, so for this set all 5000 experiments were performed over the full range.  The limit on $v_{\infty}$ is set by velocity dispersion of the Orion Nebula Cluster, $\sigma \sim$2--3 \kmss \citep{jones88}.  If the velocity dispersion in the region is Maxwellian, the integrated probability of an encounter with $v_{\infty} > 15$ \kmss is less than 1 per cent with a 1D velocity dispersion of 3 \kms.

As noted above, strictly speaking the initial conditions are not chosen according to equal probabilities. Such a sampling would be proportional  to the area of the surface element orthogonal to $v_{\infty}$, and is achieved by a uniform sampling of $b^2$ where $b$ is the impact parameter. However, the underlying physical motivation for this work is such that the stellar masses, range of interesting periastra, and likely encounter velocities place us in a gravitationally focused encounter regime, where the paths of the stars deviate substantially from rectilinear motion (see figure \ref{ICschematic}). Following the familiar derivation in \citet{binney08}, in the two-body approximation the periastron separation between the binary and the single is related to the masses, impact parameter and velocity at infinity via
\begin{equation}
b^2 = r_p \left[ r_p  + \frac{2 G (m_1+m_2+m_3)}{v_{\infty}^2} \right]
\label{focusing}
\end{equation}
The second term in brackets is associated with the gravitational focusing of the orbits. When this term is dominant, the right hand side is approximately linear in $r_p$, in which case $b^2 \propto r_p$ to a good approximation. The largest $r_p$ we use is 60 au, and the smallest value of the focusing term occurs with $v_{\infty} = 15$ \kmss in the (10,3)+10 runs, where it takes a value of approximately 181 au. Thus our entire parameter range is in a strongly focused regime, and our uniform sampling of $r_p$ is very close to the equal-probability sampling, which is formally proportional to $b^2$. We can use our setup as a reasonable proxy for one specifically tailored to determining scattering outcome cross sections, and we do this in Appendix \ref{scatteringappendix} to verify that our results are in agreement with previous work.

\subsection{Experiment termination and characterisation}
The goal of these experiments is to determine if there are regions in parameter space that can lead to the retention of disc material in a binary-single encounter that has a large final relative velocity.  In general, an encounter between a binary and a single star can result in one of two outcomes: ionisation of the system into three single stars, or an end state consisting of a binary and a single.  The latter possibility can be further divided into a flyby, in which the initially single star remains single, or an exchange in which one of the original binary members is displaced by the intruder.  For our choice of initial conditions ionisation is energetically forbidden, as the internal energy of the binary is greater than the kinetic energy associated with the binary--single orbit, i.e. $v_{\infty} < v_c$.  All of our experiments must terminate in a binary and a single.  

The interplay of three bodies in a scattering event can be quite complex and last for an unpredictable length of time \citep[e.g. figure 3 of ][]{hut93}, and some way to determine the end of a simulation is required beyond integrating for a fixed period.
During each experiment we track the size of the three stars' configuration, measured by their total moment of inertia in the center of momentum frame, 
\begin{equation*}
I_3=\sum_{i=1}^3 m_i {\bf r}_i^2.
\end{equation*}
We terminate the runs when two criteria are met. First, in the two body approximation (i.e. when the binary is treated as its center of mass) the binary--single orbit is hyperbolic. Second, 500 years have past since the last minimum in $I_3$, which we then define as the time of the scattering. At this point we have the three stars arranged as a binary and a single, each component surrounded (or not) by a cloud of disc particles. In physical units, the integrations range from about $10^3$ to $10^4$ years, roughly following a power law as $t^{-3}$ between those limits.
  
In order to determine final remnant disc membership, we rely on a circularisation scheme similar to one that has been used in previous work \citep{hall97,moeckel06}.  At the end of the simulation, the binary is treated as a single body at its center of mass.  Disc particles that are energetically bound to this center of mass have passed the first cut.  These then have their orbital angular momentum calculated, and a Keplerian orbital period determined.  In other applications, the disc particles may be placed on circular orbits corresponding to their specific angular momentum, the argument being that viscous effects and shocks between flows of material on crossing orbits will damp any eccentricity.  In this way a circularised and settled disc can be reconstructed without simulating hydrodynamic forces for long time periods. 

 Because of the specific constraints of the present problem, some care must be taken and we do not directly reconstruct the disc profile in this way.  Instead, we impose two additional cuts to determine the final disc membership. First, the orbital period of the disc particle about the binary must be less than 500 years, the time between the scattering event and the present day.  Second, the periastron of the particle's orbit around the binary center of mass must be less than 50 au, the observed size of the disc.  These dual constraints ensure that the disc particle will have passed within 50 au at least once during the 500 years available to circularise the remnant disc, and will probably have had several orbital periods to settle into a disc-like state.  We do not attempt to reconstruct the disc profile at all, but rather just track the amount of disc material that satisfies these constraints and thus estimate the mass of the disc within 50 au. 
 
 This method removes any tidal tails or wispy extended clouds from our calculation of the remnant disc mass. In some cases it excludes material that could very well be part of a the disc, either on longer timescales or in a full hydrodynamic calculation, but because of the tight time constraints of this problem we do not assign it to the remnant. Note that current observations of the disc (of both SiO
masers and radio continuum) show no signs of extended
structure at radii larger than 50 au, although future sensitive observations of thermal molecular lines and dust continuum (e.g. with ALMA) may reveal weaker emission associated
with extended gas remnants.
 
 Our approach should be conservative; including hydrodynamics, so that particles are not free to pass through the disc but would rather shock and become incorporated into the disc, would probably increase the amount of material in the disc.
  With the mass of the disc remnant interior to 50 au determined, we deem an encounter successful if it results in both a high velocity $v_{rel}$ between the binary and the single, and retains at least 10 per cent of the initial disc around the binary.  In figure \ref{examples}, we show examples of two encounters with equal mass stars, one simple (and successful) and one complex, showing the paths of the stars and the final distribution of disc particles 500 years after the encounter. The eccentric appearance of the disc in the simple case is at least partly a consequence of the dissipationless physics we are modeling; the detailed morphology of the disc remnants should not be interpreted as meaningful in this work.
  
\subsection{Integration accuracy}
The distribution of energy errors in the simulations is approximately log-normal, with the mean ${\rm log} (\Delta E / E_0)$ = -4.93 and standard deviation 0.53.  Experiments with the binary and its circumbinary disc in isolation show that the accumulated error in the integration of the disc particles and the stellar particles are of the same order of magnitude. If we were concerned about the precise details of a particular set of initial conditions, we would strive for higher accuracy.  Since we are more interested in a broad statistical survey of parameters, previous work has argued that more lenient energy tolerances are acceptable as long as conclusions don't rely on the details of a few integrations.  \citet{valtonen74} found little accuracy dependence in the statistics of a suite of approximately 200 simulations with mean relative energy errors ranging from $5\times10^{-4}$ to $3\times10^{-2}$.  \citet{pflamm-altenburg06} compared the statistical results of 1000 four-body decay experiments with mean energy errors ranging from about $6\times10^{-12}$ to $6\times10^{-2}$, and recovered the same results to within $1\sigma$.  \citet{hut83} used a relative energy error limit of $10^{-2}$, and their results compare very well to those of \citet{fregeau04}.  In Appendix \ref{scatteringappendix} we show that our experiments, although not tailored to the same goal, acceptably match the results of \citet{hut93}.  We conclude that our energy tolerance is sufficient.

\section{Results}
\label{results}
\subsection{Initial conditions resulting in success}
We begin by identifying which regions of parameter space can lead to a successful scattering event. Recall that we define success as an experiment in which a relative velocity $v_{rel}$ between the single and the binary of at least 20 \kmss is obtained, while at the same time 10 per cent of the original disc material is retained around the final binary.

\subsubsection{Results for (10,3)+10}
\begin{figure*}
 \subfigure{
 \includegraphics[scale=.72]{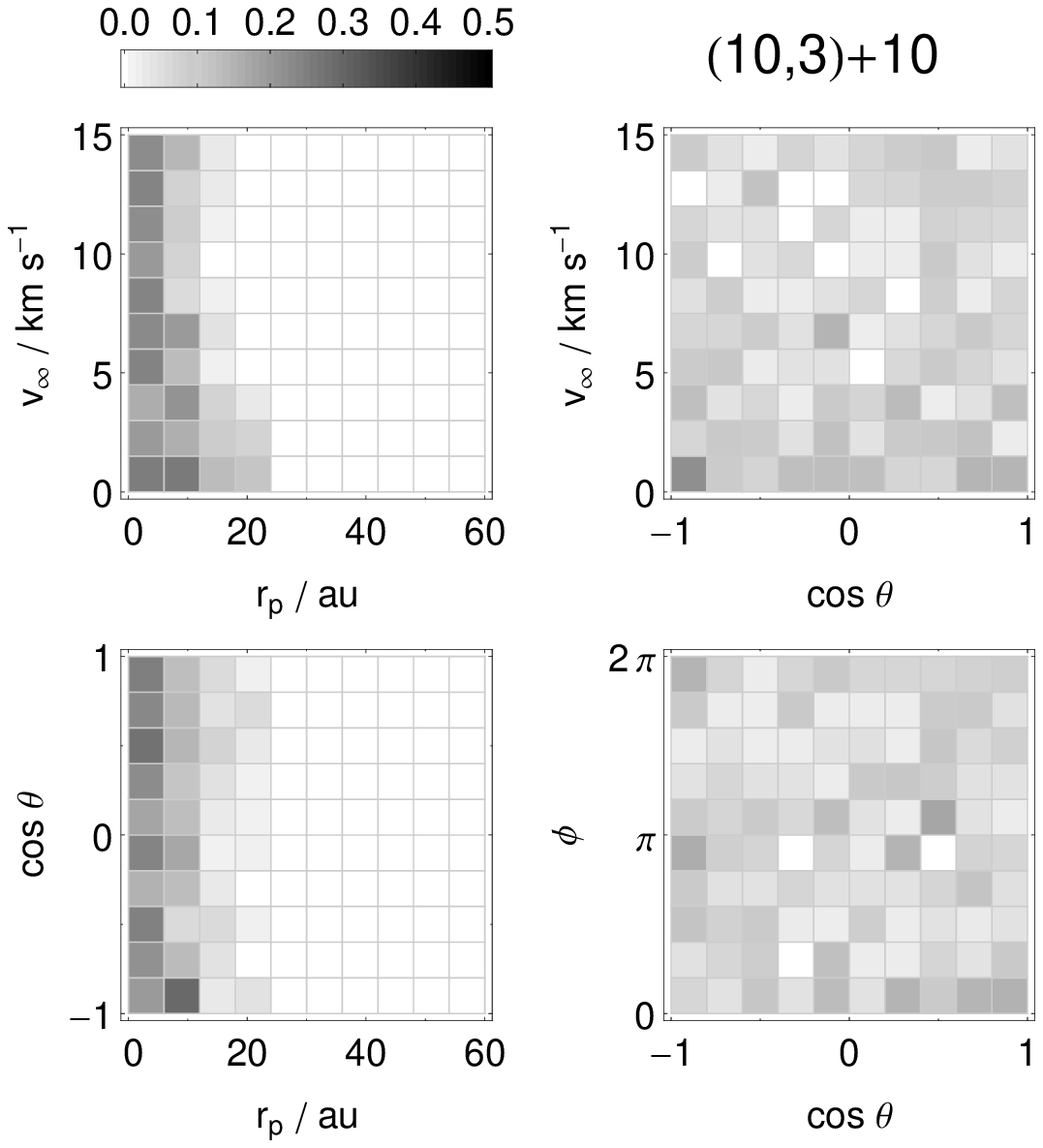}
 }
 \hspace{5mm}
\subfigure{
 \includegraphics[scale=.72]{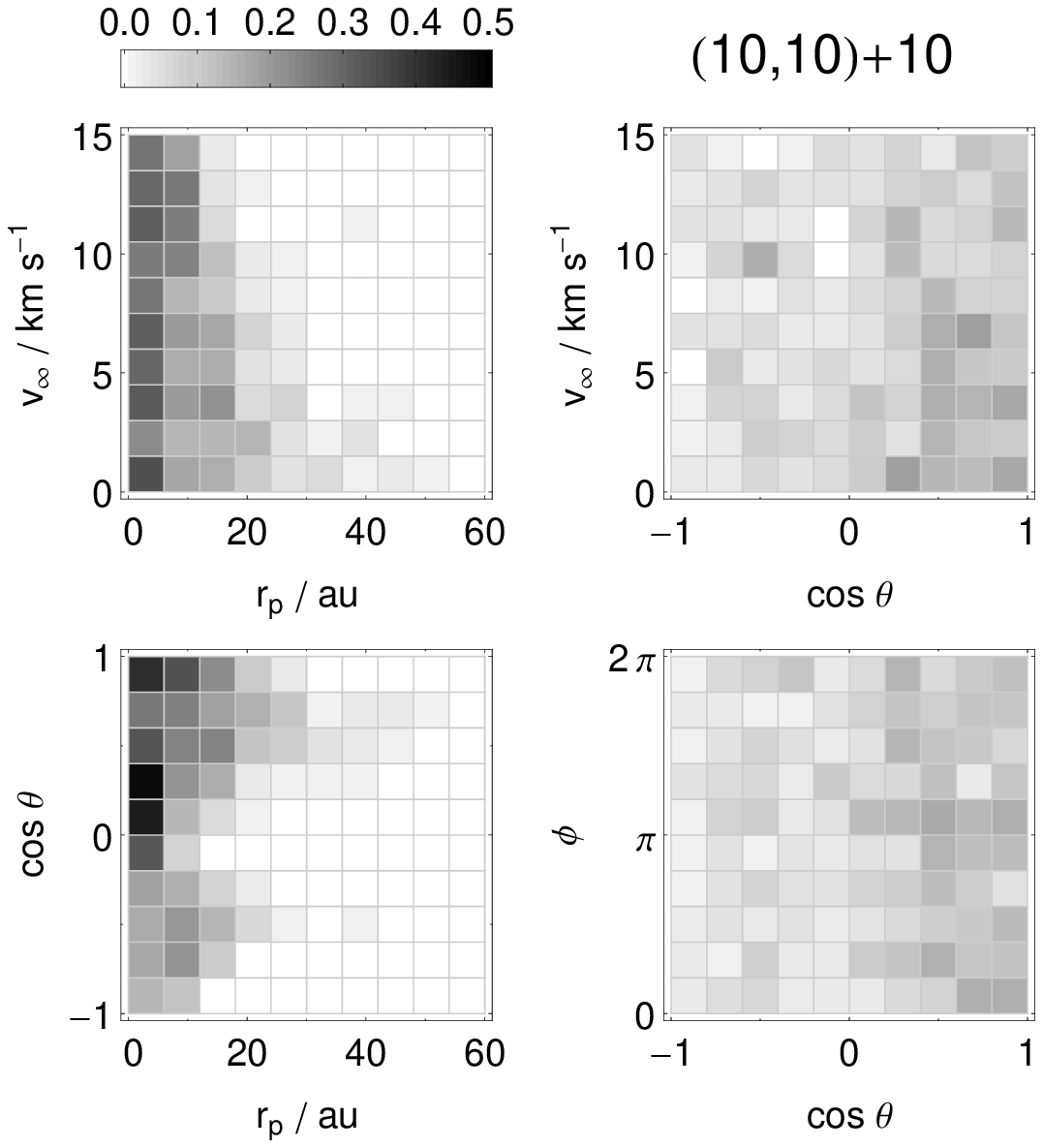}
 } 
 \caption{
 Four projections through parameter space for the (10,3)+10 case (left panels) and the (10,10)+10 runs (right panels). The graylevel shows the percentage of initial conditions resulting in both a high final relative velocity between the binary and the single, $v_{rel} > 20$ \kms, and retention of 10 per cent of the disc around the binary. In both sets, small values of the periastron separation $r_p$ are preferred for successful encounters. The equal mass runs have a higher percentage of successful encounters with prograde (cos $\theta > 0$) encounter orientations.
 }
 \label{summaries}
\end{figure*}

This set of 4500 encounters begins with an unequal mass binary composed of a 10 \msuns primary and a 3 \msuns secondary.  In the left side of figure \ref{summaries} we show four projections through the space of initial conditions, covering the most interesting combinations of parameters. In each panel the graylevel shows the percentage of runs in each bin that were successful.

Looking first at the $v_{\infty}$--$r_p$ plane, we note that all successful encounters have a periastron less than roughly 20 au.  Because of this, after completing 2000 runs we restricted the remaining 2500 to have $r_p < 30$ au. Initial conditions were drawn from the same quasi-random number generator as all the other runs, we simply did not integrate them if they missed our $r_p$ cut. The sampling density over the interesting portion of parameter space is thus identical to the (10,10)+10 set, consisting of 7000 experiments.  When constructing the histograms that follow, the small number of points with large velocities at $r_p > 30$ au are weighted according to their lower sampling density, although this is a minor effect.  There were some low-$v_{\infty}$ encounters with larger $r_p$ that resulted in high velocities, although none of them retained a disc.  These high-$r_p$ encounters all occured near \costheta$ = 1$, a nearly co-planar prograde encounter.  If the phase of the binary orbit is favorable, this configuration can maximize the interaction time between the intruder and the secondary.  The panels on the right show that there is no preferred combination of $v_{\infty}$ and \costhetas leading to successful encounters, and no preferred orientation of the intruder's initial orbit.  

\begin{figure*}
 \subfigure{
 \includegraphics[scale=.58]{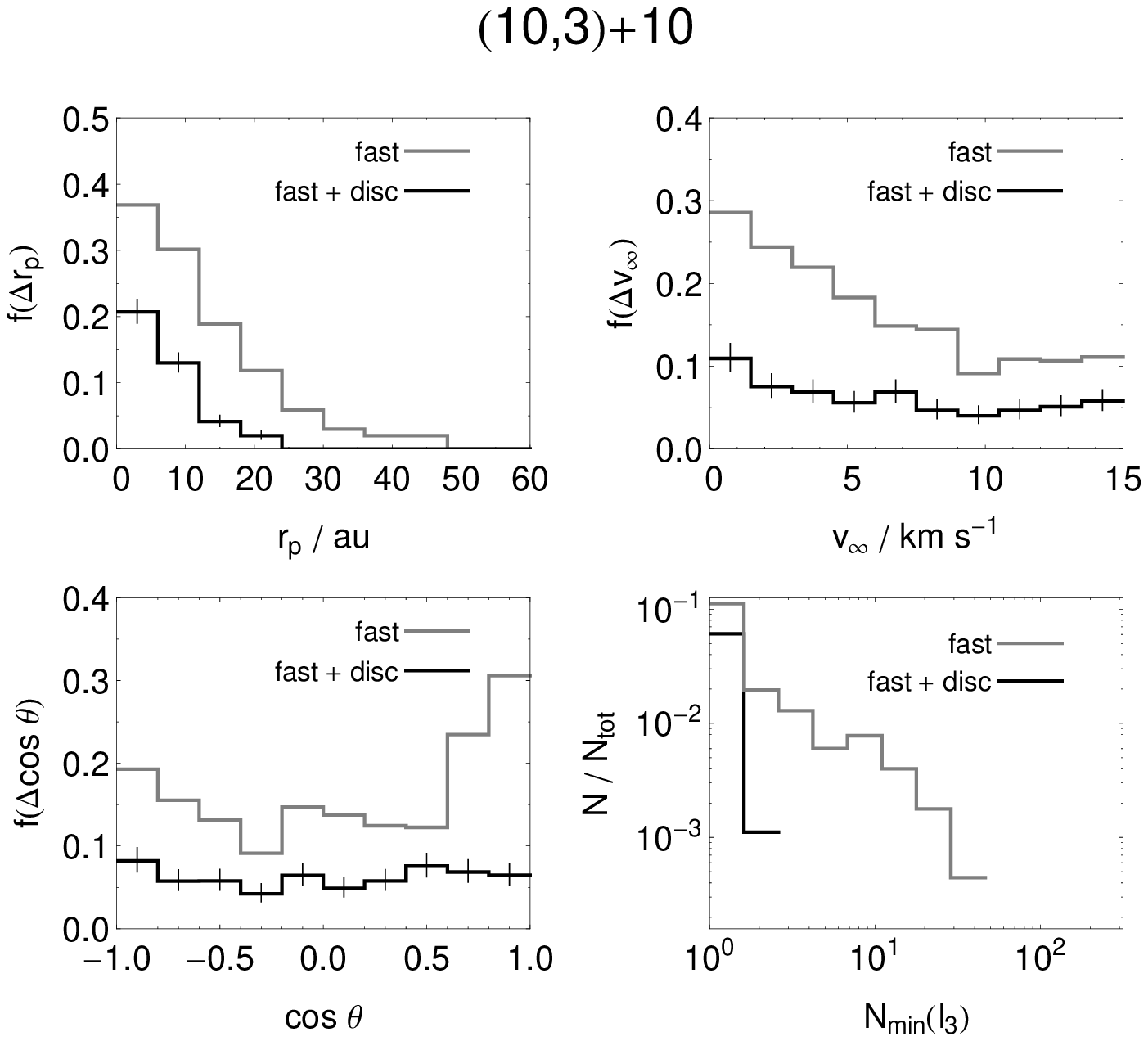}
 }
 \hspace{3mm}
\subfigure{
 \includegraphics[scale=.58]{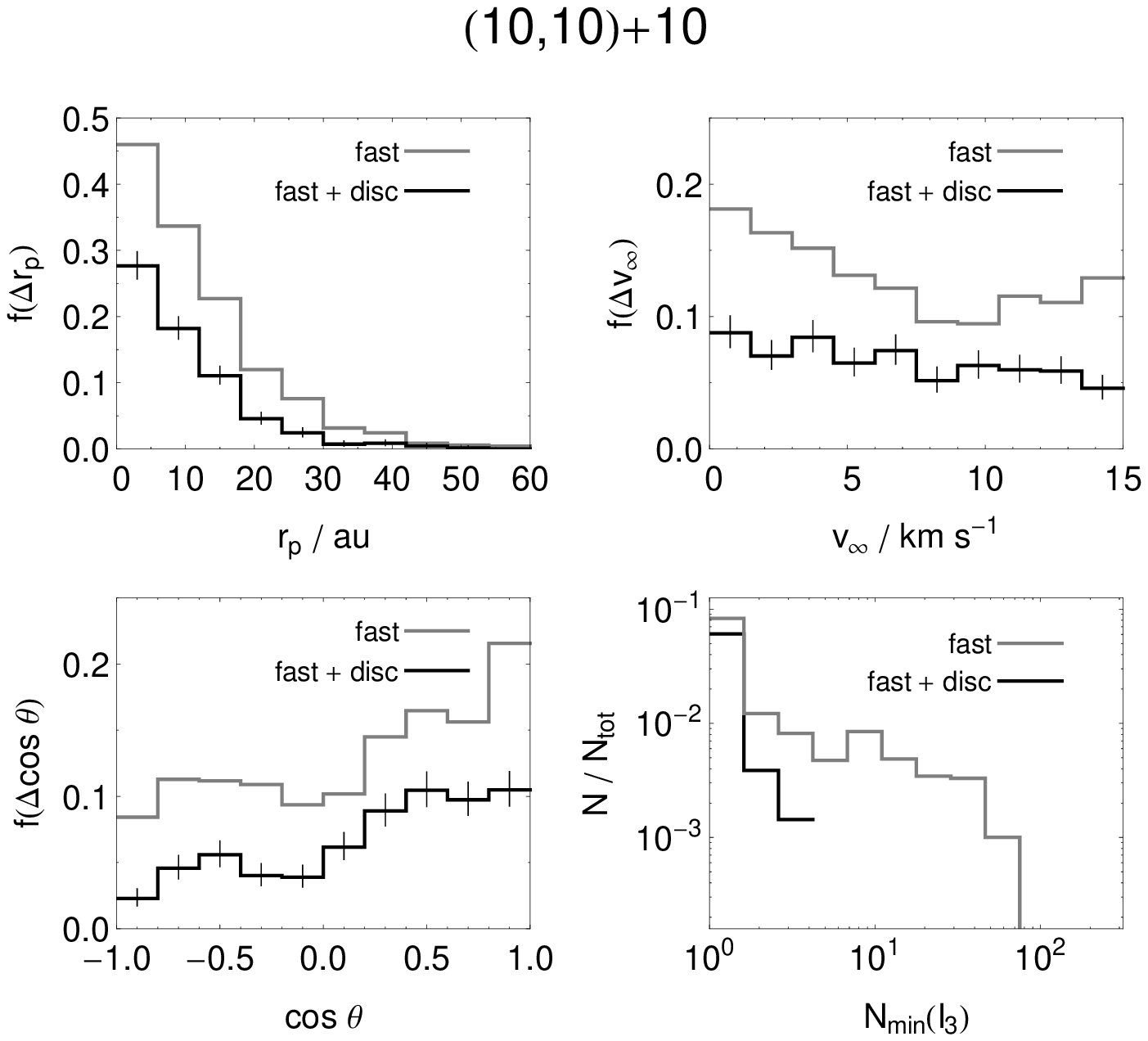}
 } 
 \caption{Histograms for the (10,3)+10 runs (left panels) and (10,10)+10 runs (right panels) showing the fraction of runs that resulted in $v_{rel} > 20$ (``fast"), and the fraction that also retained 10 per cent of the initial disc (``fast+disc") as a function of $r_p$, $v_{\infty}$, cos $\theta$, and N$_{min}(I_3)$. 
The last is the number of minima in the three-body moment of inertia $I_3$ showing the distributions of encounter complexity for those same sets of runs.}
 \label{histograms}
\end{figure*}

In the left side of figure \ref{histograms} we show the marginal distributions of the encounter results for some of the parameters (i.e. all other parameters have been integrated over). In three of the panels we plot histograms showing the fraction $f$ of encounters with high final $v_{rel}$ (`fast') and successful encounters (`fast+disc') for bins of $r_p$, $v_{\infty}$, and \costheta. The fast+disc histograms have Poisson 1$\sigma$ error bars. There is a clear tendency for low  $v_{\infty}$ cases resulting in more fast results, but this results in only a mild dependence for the fast+disc runs, concentrated in the lowest $v_{\infty}$ bin. There is likewise structure in the \costhetas histogram for the fast cases, but the fast+disc histogram is virtually featureless. In contrast, there is an obvious dependence on $r_p$, with a dramatic falloff in successful encounters over the range $0 < r_p < 25$. Small-$r_p$ encounters are not only more likely to result in a high ejection velocity, but the fraction of those fast encounters that retain a disc likewise increases with smaller $r_p$.

\subsubsection{Results for (10,10)+10}
In this set of  7000 encounters all three stars have the same mass, 10 \msun.  We again begin by looking at four projections through the initial conditions, shown on the right side of figure \ref{summaries}.  In contrast to the (10,3)+10 case there are encounters at large periastron resulting in success, those occurring at low $v_{\infty}$ and prograde inclinations.  Besides the increased range of $r_p$ that allows disc retention and the somewhat higher percentages at low $r_p$, the most striking difference between these runs and the case with a 3 \msuns secondary is the angular dependence of successful encounters.  The density of successful initial conditions increases significantly with prograde encounters ($0 <$\costheta$<1$) relative to retrograde cases.

On the right side of figure \ref{histograms} we show the marginal distribution histograms for the equal mass encounters.  The general trend of the $r_p$ dependence is the same as the (10,3)+10 case, although with a broader range.  While there is some bias toward low-$v_{\infty}$ cases resulting in high final velocities, this trend is again less apparent in the fast+disc cases.  The \costhetas histogram contrasts sharply to the (10,3)+10 experiments.  At retrograde inclinations the fraction of successful runs is fairly flat, and begins to rise as the encounters become prograde. 

\begin{figure}
 \includegraphics[width=84mm]{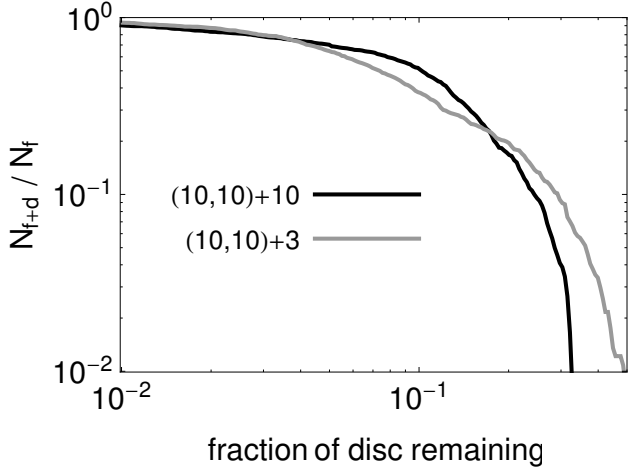}
 \caption{The percent of runs with high velocities ($N_f$) that also retain a disc ($N_{f+d}$) for different fractional disc retention criteria. The results quoted in the bulk of the paper are for 10 per cent disc retention.}
 \label{discmassdependence}
\end{figure}

\subsubsection{Dependence on the remnant disc mass cutoff}
Our choice of 10 per cent of the disc remaining bound as a criterion for success is somewhat arbitrary. Given the lack of constraints on both the initial and current disc masses, this is in some sense unavoidable. This choice does have some effect on the results. In figure \ref{discmassdependence} we show the fraction of encounters with $v_{rel}>20$ \kmss that retain discs of different masses. As the cutoff on the remaining disc mass increases fewer runs result in success, with a precipitous drop-off above about 20 per cent of the disc remaining, especially for the equal-mass runs. 

To a good approximation, the behavior of the results shown in figures \ref{summaries} and \ref{histograms} scale with these curves. Versions of these figures with cutoffs of 2.5 and 20 per cent rather than 10 per cent display very little functional variation in the initial conditions leading to successful outcomes, and a simple renormalisation suffices for the purposes of this paper. Thus our quoted results with the 10 per cent criterion can be thought of as representative of results up to a requirement of $\sim 20$ per cent disc retention, modulo a scaling of order unity; above this, the order of magnitude of the results changes and success becomes much less likely.

\subsection{Disc survival with encounter complexity}
Intuitively, one would expect a complex encounter (such as that shown in the lower row of figure \ref{examples}) to disrupt a disc much more efficiently than a relatively clean and fast exchange.  This intuition is borne out by the experimental results.  In the lower-right panels of figure \ref{histograms} we show the number of close triple encounters, measured by the number of minima $N_{min}$ in $I_3$, for all the experiments performed,  for the high-$v_{rel}$ runs, and for successful encounters. These plots are normalized to the total number of experiments performed for each set. The distribution of $N_{min}$ is quite broad, peaked at low values but extending from 10s to 100s of minima. 

The fraction of all runs that are classified as fast is fairly independent of $N_{min}$, at around 10 per cent of all encounters. In contrast, the encounters that leave a remnant disc are almost all single-minimum events.  For the (10,3)+10 case less than 3 per cent of successful encounters go through multiple close triple passages, and for (10,10)+10 runs less than 6 per cent do. This is the most restricting component of a successful scattering event; it must be a clean encounter, either exchanging the intruder for a binary member or ejecting the intruder with a minimal amount of interaction.

\subsection{Final relative velocities and binary orbits}
\begin{figure}
 \subfigure{
 \includegraphics[scale=.44]{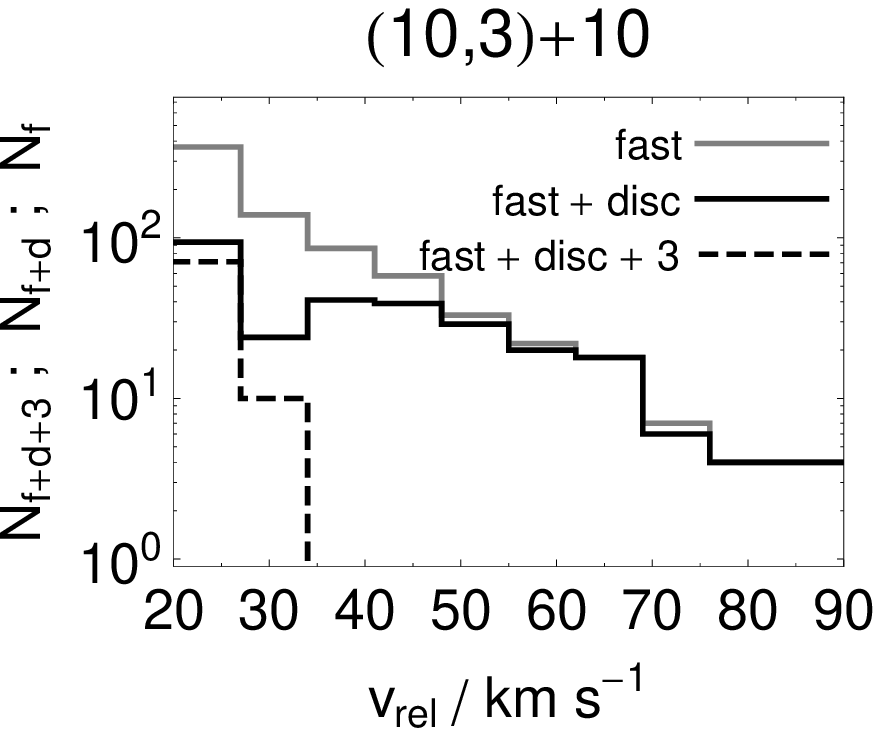}
 }
\subfigure{
 \includegraphics[scale=.44]{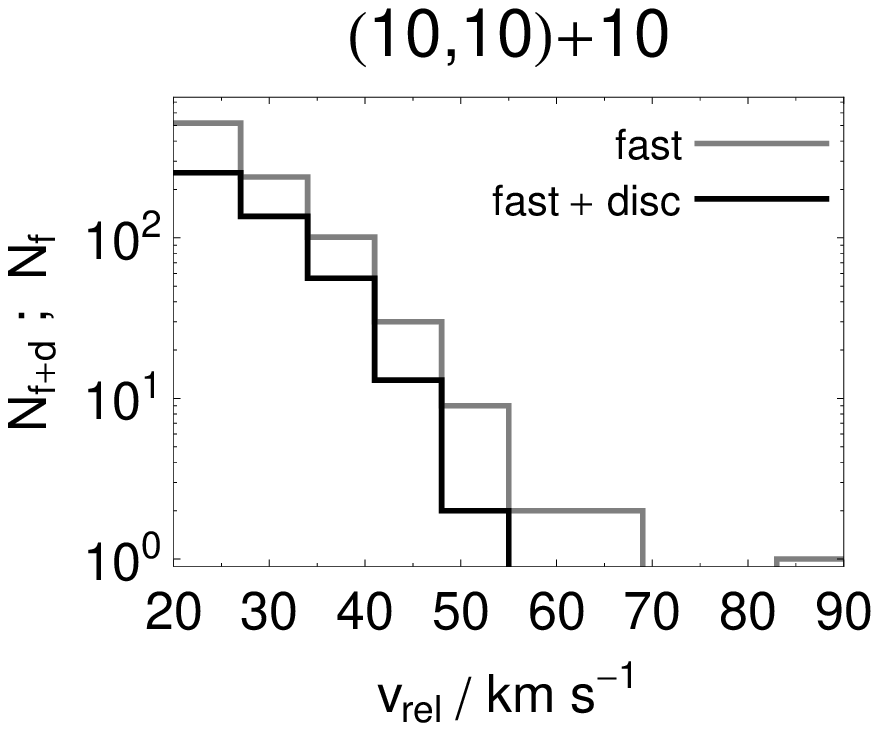}
 } 
 \caption{The distribution of final velocities $v_{rel}$ for the (10,3)+10 case ({\it left}) and the (10,10)+10 case ({\it right}).  Shown are the runs that resulted in $v_{rel} > 20$ \kmss (``fast"), the subset that also retained 10\% of the initial disc (``fast+disc"), and for the (10,3)+10 runs the subset that retained the lightest star as a member of the final binary (``fast+disc+3").  For the (10,3)+10 plot, all cases with $v_{rel}>85$ \kmss are represented in the final bin.}
 \label{finalvelocities}
\end{figure}

\begin{figure*}
 \subfigure{
 \includegraphics[scale=.8]{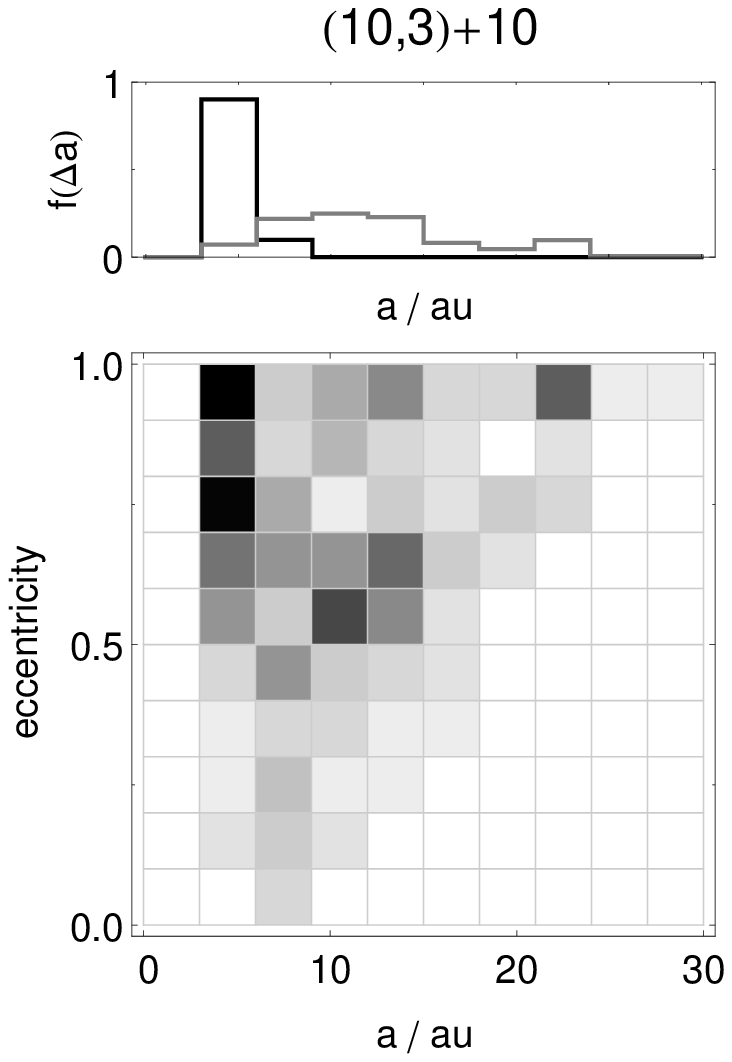}
 }
 \hspace{5mm}
\subfigure{
 \includegraphics[scale=.8]{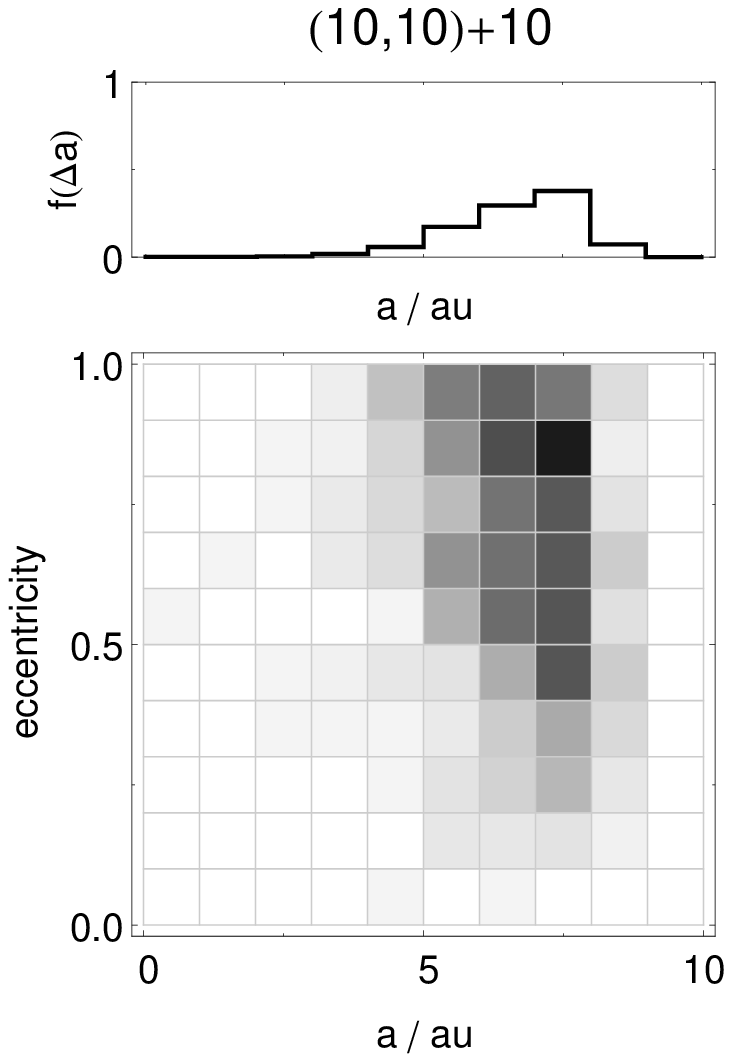}
 } 
 \caption{The location of the final binaries in the semi-major axis--eccentricity plane for all cases with $v_{rel}>20$ \kms.  Graylevels show the density of points in each cell. The histograms show the distribution integrated along the eccentricity, i.e. showing the fraction of runs in bins of $a$ only. For the (10,3)+10 runs, the cases that retain the 3 \msuns star are shown in the black histrogram, while those where it is ejected  are shown in grey.}
 \label{finalbinaries}
\end{figure*}

In figure \ref{finalvelocities} we show the distribution of the final relative velocities $v_{rel}$ for both sets of runs.  There are several previous results concerning the escape velocity of the decay of three equal-mass bodies from an initially binary-free setup, e.g. \citet{mikkola86} and \citet{sterzik98}, with good agreement to an analytic approximation generalized from \citet{heggie75}.  These distributions rise sharply to a maximum and decay over a longer range of velocities.  While not directly applicable to our study, the general behavior of our equal-mass results mirrors the high-velocity tail of those earlier works.  The (10,3)+10 results show more structure than the equal-mass case.  Here we plot both the distribution of all encounters that retain a disc, as well as those that keep the 3 \msuns star in the binary.  These cases are exclusively below $v_{rel}=35$ \kms, with all the higher velocity cases resulting from scatterings where the 3 \msuns star is ejected. At higher relative velocities the fast and fast+disc populations are nearly identical. This is due to momentum conservation in the three-body scattering. These encounters are all ones in which the 3 \msuns star is swapped for a 10 \msuns star and the recoil velocity of the now more massive binary is much less than that of the liberated low-mass star. A low binary velocity combined with a deeper potential makes disc retention easier.

The velocity behavior is reflected in the orbit of the binary after the scattering, which we show in figure \ref{finalbinaries}.  The relative velocity between the binary and the single at the end of the scattering experiment is a function of the change in energy of the binary as a result of the encounter, and any increase in $v_{rel}$ compared to the initial $v_{\infty}$ must be compensated by an increase in the binary's binding energy.  For the equal mass cases considered in the (10,10)+10 runs the energy change is simply inversely proportional to the binary's semi-major axis; those runs with $v_{rel} > 20$ \kmss leave binaries with a semi-major axis of $6.7\pm1.0$ au and eccentricity $0.66\pm0.22$. For simplicity we have characterised the distributions by their means and standard deviations, although the eccentricity distributions in particular are quite broad. While these equal mass stars are effectively identical in these simulations, we note that in 51 percent of the encounters with $v_{rel} > 20$ \kmss an exchange occurs, and the initially single star takes the place of one of the original binary members.

With the unequal masses in the (10,3)+10 runs, the final binary's semi-major axis depends on which star is ejected from the system.  When the 3 \msuns star is retained in the binary, the semi-major axis must be reduced in order to boost the relative velocity; the semi-major axis distribution for these runs is characterized by a mean value of $5.2\pm0.77$ au, with eccentricities $0.78\pm0.18$.   When the 3 \msuns star is ejected, however, the semi-major axis of the now more-massive binary can be larger than the initial value.  The semi-major axis distribution for these runs is much broader than the other cases.  There is a main peak around $12.6\pm5.6$ au and eccentricities $0.67\pm0.24$, but there is also a population above 20 au with high eccentricity. This population is problematic for the survival of a roughly co-planar remnant disc of 50 au.

 The high eccentricity of all the encounters leading to high velocities is notable; if Source I is found to be a circular binary, it seems unlikely that it took part in a recent scattering event. Note that the binarity of Source I is not observationally established, as length scales $<10$ au are beyond the resolution of current instruments.

\subsection{Remnant disc orientation}
\label{discorientationsection}
\begin{figure*}
 \subfigure{
 \includegraphics[scale=.56]{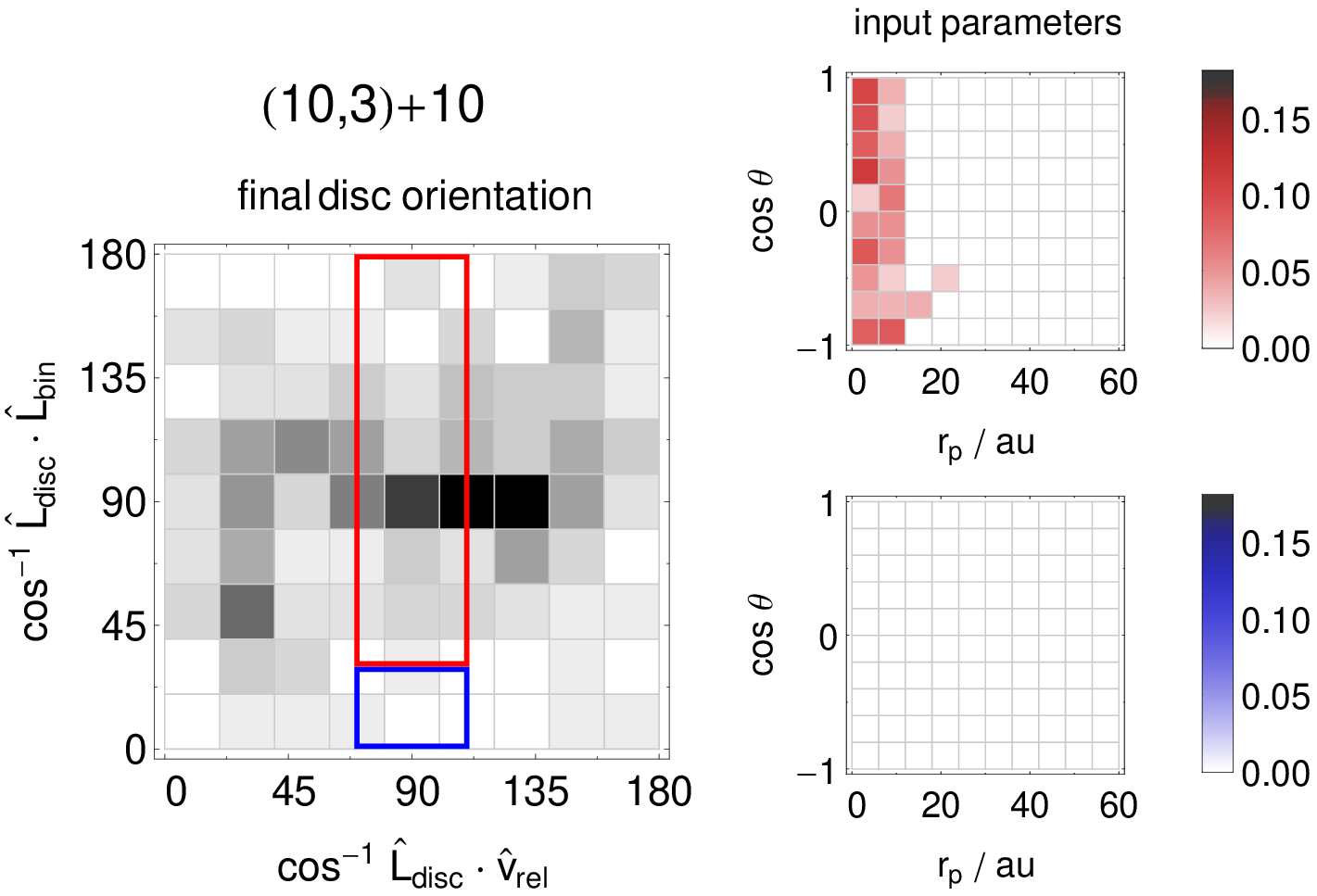}
 }
 \hspace{3mm}
\subfigure{
 \includegraphics[scale=.56]{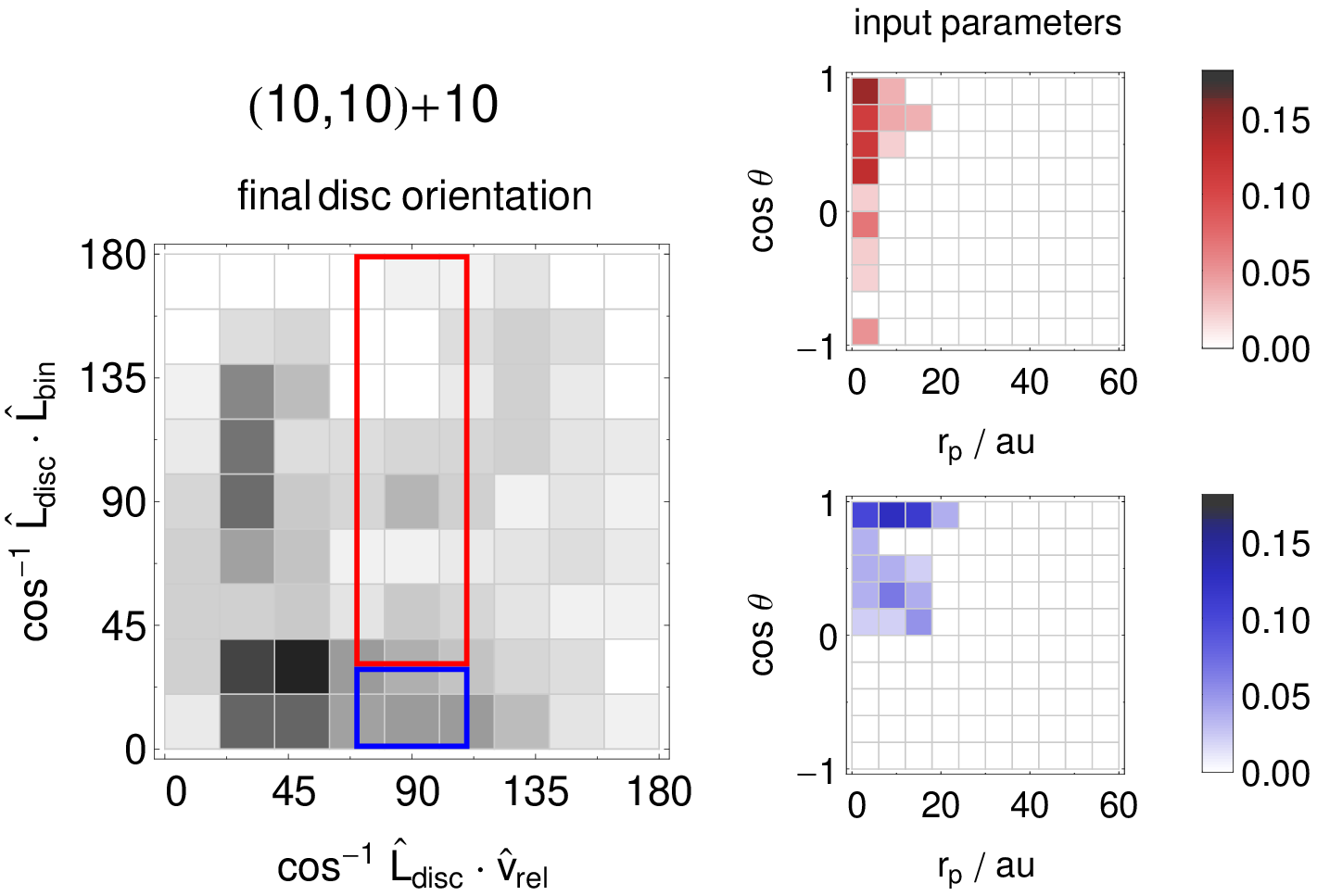}
 } 
 \caption{
The large panel of each set of plots shows the distribution of successful encounters in the plane of two disc orientation angles; the angle between the remnant disc orientation and the binary's orbit, and the angle between the disc orientation and the binary's velocity. For these purposes the disc and binary orientations are taken as their respective angular momentum vectors. Observations have shown that the disc symmetry axis and the proper motion of Source I are roughly orthogonal (see Figure 1). We have overlaid boxes around the region of this plane where these vectors are within 20 degrees of orthogonality. The blue box encompasses results where the binary and disc angular momenta are within 30 degrees of alignment, and the red box encompasses all other binary--disc alignments. The smaller plots trace these results back to their input parameters, showing the density of initial conditions in the $r_p$--cos $\theta$ plane that lead to final disc orientations in the red and blue regions of interest.
}
 \label{angles}
\end{figure*}

A final observable in the BN/KL system that we can compare our simulations to is the relative orientation of the disc and the presumed binary's velocity vector. The proper motion of Source I is from north-west to south-east, and the nearly edge-on disc has a spin axis aligned north-east--south-west \citep[][see figure \ref{propermotions}]{goddi11}. We can search our data for experiments that yield this roughly orthogonal alignment.  In figure \ref{angles}, we show for both sets of simulations the relationship between two angles for all successful runs: the angle between the disc and the binary angular momentum, cos$^{-1}({\rm {\bf \hat{L}_{disc}}} \cdot {\rm {\bf \hat{L}_{bin}}})$,  and the angle between the disc and the binary--single relative velocity, cos$^{-1}({\rm {\bf \hat{L}_{disc}}} \cdot {\rm {\bf \hat{v}_{rel}}})$.  The disc angular momentum is taken to be the sum of the angular momenta of all disc particles that are deemed part of the final disc remnant, calculated in the reference frame of the binary. While each set has some higher-density regions in this plane, in general most combinations of these angles are represented.

We have highlighted all runs that have a mostly orthogonal orientation between the disc angular momentum and $v_{rel}$, which we take as within 20 degrees of orthogonal, with boxes. The blue box encompasses all nearly-orthogonal runs that also have the disc and the binary aligned to within 30 degrees; the red box encompasses all other disc--binary orientations. In the smaller panels of the figure, we trace back to the initial conditions to see if these interesting boxes are populated from a specific region of parameter space, plotting the density of runs that end up populating each box. The only interesting features turns out to be in the $r_p$--cos $\theta$ plane.
The first notable result is that the (10,3)+10 series contains no runs that led to alignment between the binary and a disc that is orthogonal to the relative velocity between the single and the binary. The equal mass series, in contrast, has several runs in this category. Crucially, the initial conditions these runs came from are much smaller than the total parameter range we covered. Roughly 10 per cent of all initial conditions with periastra less than about 20 au and prograde encounter angles result in high stellar velocities and remnant discs with similar orientations to Source I.

\subsection{Likely effect of relaxing assumptions}
All of our simulations have taken the binary as initially circular. We can draw on extensive binary--single scattering work to estimate the effect of this simplifying assumption. \citet{hut93} presents scattering results for equal-mass systems at three eccentricities: $e=0$, 0.7, and 0.99. He shows that the cross sections for scattering results (see the appendix for some discussion of this) depend only weakly on the eccentricity; increasing the eccentricity from 0 to 0.7 approximately doubles the cross section for both simple exchange encounters and resonant encounters, and a further increase to 0.99 has minimal effect. This increase in encounter cross section is mainly manifested in encounters that only slightly change the binary energy, however \citep{hut84}. Over the range of velocities we consider, the cross section for a given energy change to the binary as a result of a scattering encounter in the range of interest, i.e. those encounters that generate $v_{rel}>20$ \kms, are almost identical for all three eccentricities. The total number of sufficiently velocity-boosting encounters is thus largely insensitive to the eccentricity of the binary, and our zero-eccentricity assumption should not affect these results. 

We also assume that the binary and its disc are largely dynamically unprocessed, in the sense that the disc is coplanar to the binary. Our results do rely on this assumption; if the disc and the binary are seriously misaligned prior to the encounter, our effort to trace back the interesting outcomes to their initial conditions (figure \ref{angles}) is compromised. While there would almost certainly be a set of encounters that are more likely to lead to the proper disc--$v_{rel}$ orientation, it might not be as nicely contained as in the coplanar case. 

The initial parameters of our disc, for which we have attempted to choose plausible values, are also an assumption. If the original disc is considerably larger or more compact than our choice of 100 au, the fractional amount of disc material retained will be different than the values quoted here. Because the bulk of the retained material is originally in the inner regions of the disc, the conclusion that non-negligible amounts of material can remain bound should hold true as long as the inner regions of the disc are still intact. The effect of tidal or direct interactions with low mass cluster members prior to the encounter with the BN object should not greatly alter these results either, provided they do not get captured into a wide binary/hierarchical triple; although in this case the disc would eventually be fully disrupted \citep{moeckel06}, and this seems incompatible with the observed system.

Finally, we are working under the assumption that the gravitational interaction between the stars and the disc is the dominant physical mechanism at play. We are in effect simulating the dense midplane of the disc, rather than the surface or atmosphere that is actually observed in radio continuum and SiO masers. Our assumption is thus that majority of the disc mass is in the molecular gas of the disc midplane, obscured from observations. If the disc were totally ionised, ignoring the hydrodynamic forces would certainly not be a valid approximation. Other effects that may work to sculpt the disk morphology, such as photoevaporation, could be at work around a hot source like Source I \citep{hollenbach94}. Indeed, with the nominal remnant disc mass in our simulations of $\sim0.01$ \msun, a photoevaporative wind with $\dot{M}\sim10^{-5}$ \msuns$yr^{-1}$ could act rapidly enough to be important over 500 yr. We have worked mostly with the fractional disc mass rather than its absolute value, however; increasing our initial disc mass by an order of magnitude, which would still be quite small compared to the binary mass, would make the orbital timescale the only relevant one over 500 yr. While we regard the `gravity as the dominant force' approximation to be a reasonable one for shaping the size of the remnant disc, further work with more massive discs (including at least hydrodynamics) is necessary to confirm this.

\section{Discussion}
\label{discussion}
Disc survival in encounters involving more than two stars is an unexplored problem. The combination of star-disc interactions and three body scattering is quite relevant in star formation theories where stellar dynamics play a dominant (or at least perturbing) role in the accretion process \citep[e.g.][]{bonnell01a,peters10}. 
It is notable that both Orion BN/KL and Cepheus A HW2 \citep{cunningham09}, two of the closest and best studied sites of ongoing massive star formation, show evidence of stellar interactions during the accretional growth of a massive YSO. The parameter space of binary-single scattering with discs present is huge, and the strong evidence of a multiple encounter in the BN/KL region has provided a physically motivated point to focus our initial investigations around. 
Our simulations described in Section~\ref{results} enable us to constrain the likely encounter parameters that could lead to the observed BN and Source I velocities and disc configuration. 

The first point we make is that a scenario with all three bodies of similar mass seems more likely than one with a low mass object. The low-mass star is more likely to be the ejected member of a scattering encounter; in our experiments, about 80 percent of encounters resulting in high velocities end up with the two most massive stars in the binary. If these mass ratios are similar to reality, this would imply that the BN object is likely the lowest mass star in the trio, which is incompatible with luminosity-based mass estimates of around 8-12 \msuns  \citep{scoville83,rodriguez05a}. No similar estimate of Source I's mass exists, as it is undetected in the mid-IR \citep{greenhill04}. However, based on the motion of the SiO masers and the ionising flux derived from radio continuum emission, it is estimated to be greater than 7-10 \msuns \citep[see][for a full discussion of Source I's mass]{goddi11}.  It is possible that the scattering encounter was one of the 20 percent that leaves the 3 \msuns star in the binary. In this case, the overall magnitude of the relative velocity between Source I and BN could be consistent with the observed value, and the final binary would be guaranteed to have a smaller semi-major axis after the encounter than before. However, momentum conservation and the observed proper motions suggest a mass ratio of Source I to BN of approximately a factor of two (assuming that the center of mass of the three stars is stationary in the Orion rest frame).
The remnant disc is a further line of evidence arguing against a low mass member of the system. While retaining disc material is not any more of an issue for the (10,3)+10 cases than the equal mass runs, the orientation of the disc relative to the proper motion vector is problematic. Not a single run ended up generating the observed angles, as discussed in section \ref{discorientationsection}. 
We thus turn our attention to the equal mass case.

Referring to figure \ref{histograms}, the total percentage of encounters that result in both high final velocities and a disc is not large ($\lesssim$10 percent). Importantly, though, $\sim$50 percent of those (`fast') encounters also retain 10 percent of the initial disc (`fast+disc'). 
Interestingly, the distribution of successful encounters is heavily biased toward small encounter distances; almost 30 percent of encounters with $r_p < 5$ au result in success. This is somewhat at odds with the intuitive picture many would have, that the current disc size marks the closest approach of the intruder, as this truncation mechanism has been noted and investigated in many numerical studies of star-disc encounters \citep[e.g.][]{clarke93,heller95,hall96,boffin98,moeckel06}. 
If the high proper motions were not present, truncation would in fact be an attractive explanation for the disc size. However, as seen in figure \ref{histograms}, encounters well within the current size of the disc are necessary to achieve binary hardening and the consequent relative velocity boost. In fact, no encounters outside 50 au, the observed size of the disc, result in velocity boosts. 

The key to reconciling very close encounters and disc survival lies in the distribution of encounter complexities. All successful encounters are over very quickly, mostly with a single minimum in the spatial configuration of the bodies. The picture to have in mind is a clean interaction taking place within the circumbinary cavity, essentially a head on encounter between the binary and the intruder, that sends the stars on their new paths with minimal direct disruption of the disc. Thus rather than the surviving disc being a truncated remnant after the binary and the intruder undergo a Keplerian encounter, as in previous work with star-disc encounters, it consists of the material that is dragged along as the binary's trajectory is almost impulsively altered.

This scenario has several advantages compared to the idea of accreting the disc from the ISM after the encounter. While there is some evidence that small short-lived discs can be set up around an accreting object moving through material with some momentum gradient perpendicular to its velocity \citep{ruffert97,ruffert99}, the disclike structures in those simulations are smaller as a fraction of the gravitational radius ($r_d < 0.1 r_G$) than the one around Source I (with $r_G = 2 G M / v^2 \sim 250$ au, compared to the 50 au disc). Furthermore, the time between the scattering event and now is only $\sim$500 years and the distance traveled is only $\sim$700~AU, i.e. a few times the gravitational radius. Setting up an identifiable Bondi-Hoyle type flow over that short of a time and distance may be difficult. In the picture presented here, the material starts out bound and in a disc, and immediately after the encounter the remnant is still largely co-planar and bound to the binary. Over the $\sim 500$ years since the encounter, this material only has to reorganize itself from the eccentric orbits we see in our simulations into the disc observed today rather than build itself from scratch from the ISM.

In summary, our numerical work has enabled us to constrain the parameters of a stellar scattering that probably occurred in BN/KL: roughly equal mass stars, a prograde encounter between the binary orbit and the orbit of the binary-single system, and a periastron comparable to the binary semi-major axis. 
We note that this work includes only the most dominant force, gravity. This choice provides the most computationally efficient strategy for large parameter searches. We cannot, however, say anything about the circularisation of the remnant disc without including at least hydrodynamics. Additionally, magnetic fields may play a significant role in generating the observed outflows \citep{bally11}, as well as providing some support to the disc around Source I \citep[see the discussion of this point in][]{goddi11}. 
Despite these limitations, this work has significantly narrowed the range of parameters for future studies which will incorporate more complicated and computationally expensive physics. 

\section{Conclusions and future work}
\label{conclusions}
We have shown that the gravitational dynamics of a velocity-boosting three-body scattering event between an existing binary with a circumbinary disc and a single star do not preclude the survival of some fraction of the disc. Despite the chaos of the three-body encounter and the nearly impulsive velocity kick the binary center of mass experiences, there are regions of encounter parameter space that permit at least 10 percent of the disc to remain associated with the binary. The encounter type that is most permissive of disc survival is a nearly head-on one, so that the stellar dynamics occurs mainly in the cavity of the circumbinary disc. Encounters that are complex with a long-lived stellar interaction will more efficiently disrupt the disc. When we take into account the present day orientation of the disc around Source I, we are able to further constrain the encounter to one with roughly equal-mass stars, and with a prograde orientation between the orbit of the binary and that of the binary--single orbit.

We stress that this work is the first step in a systematic investigation of the dynamics at work in BN/KL, hence the conclusions must be interpreted with the first-order approach we took in mind. As we have noted throughout the paper, additional physics such as hydrodynamic, magnetic, and radiative forces will all be important to fully understand the BN/KL region. Looking forward, we cannot speculate too far about the path our explorations will take, as each addition of physics will build upon the results of the previous work. The next step is clear, however, and that is to include hydrodynamics and study the settling of the disc in a self-consistent fashion, drawing initial conditions from the restricted set that we have identified here. This will address the greatest shortcoming of this study, which is the lack of dissipation in the disc. Given the 500 year time constraint of this problem, this is an important issue to investigate for this scenario, and is the subject of ongoing work using an {\it n}-body/SPH code.

\section*{Acknowledgments}
We thank the referee for a constructive report. NM thanks the midwives of Addenbrooke's Hospital for caring for his son, Isaac Hector, born just this paper was submitted. This work was performed in part using the Darwin Supercomputer of the University of Cambridge High Performance Computing Service (http://www.hpc.cam.ac.uk/), provided by Dell Inc. using Strategic Research Infrastructure Funding from the Higher Education Funding Council for England. 

\bibliographystyle{mn2e}

\appendix
\section{Consistency check via cross sections}
\label{scatteringappendix}
As a point of comparison between our simulations and previous work, we estimate the total cross section for exchange and resonance in the equal-mass case (10,10)+10. Cross sections are customarily presented with $v_{\infty}$ measured in terms of the critical velocity $v_c$, which is where the total energy of the binary-single system is zero (see equation \ref{vcrit}). For our setup, recall that $v_c$ in physical units is 36.5 \kms.  As our limits on $v_{\infty}$ rule out ionisation of the binary, all of our simulations are in the regime $v_{\infty} < v_c$.  While this rules out comparison with the more recent tests in \citet{fregeau04}, figure 3a of \citet{hut93} includes the case of an equal mass, zero eccentricity binary scattering with a star of matching mass over a velocity regime coincident with ours. 

As a benchmark we use the data from that figure, and for consistency with that paper we define an exchange as an encounter with fewer than four minima in $I_3$ after which one member of the binary has been replaced by the intruder, and all encounters with more than four minima are deemed to be in resonance.  In studies designed to study cross sections, rather than sampling the initial conditions evenly over $r_p$, one samples over the impact parameter $b$ according to its probability, i.e. uniformly in $b^2$.  The total cross section for outcome $X$ is estimated simply by
\begin{equation}
\sigma_{X}(v_{\infty}) = \pi b_{max}^2(v_{\infty})\frac{N_{X}(v_{\infty})}{N_{tot}(v_{\infty})}.
\label{cross_section_formula}
\end{equation}
$N_{X}$ is the number of experiments resulting in the outcome of interest, and $N_{tot}$ is the total number of experiments performed out to a maximum impact parameter $b_{max}$. While we have chosen to evenly sample $r_p$ instead, in the regime where $v_{\infty} < v_c$ the gravitational focusing term of the encounter cross section dominates, and $r_p \propto b^2$ to a good approximation (see the discussion in section \ref{parameterranges}). We thus use equation \ref{cross_section_formula} directly to calculate cross sections, after binning our data logarithmically in $v_{\infty}$. The errors associated with this type of Monte Carlo estimate are fully discussed in \citet{hut83}; we show the statistical uncertainty
\begin{equation}
\Delta \sigma_{X}(v_{\infty}) = \sigma_{X}(v_{\infty}) / \sqrt{N_{X}(v_{\infty})}.
\end{equation}

\begin{figure}
 \includegraphics[width=84mm]{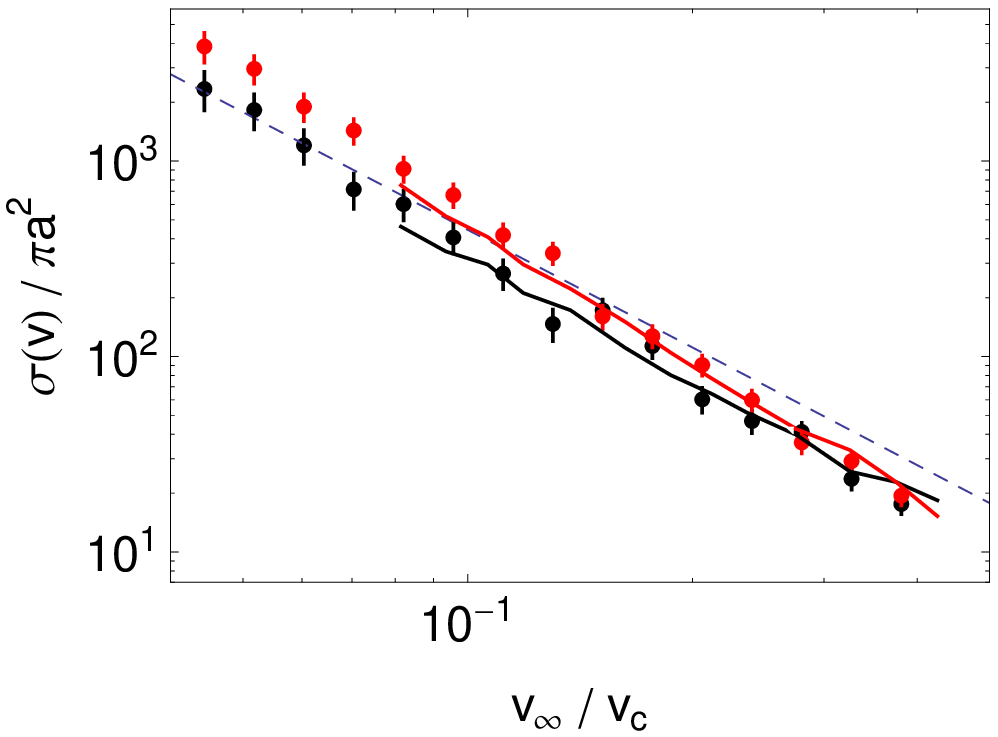}
 \caption{Cross section for direct exchange from our (10,10)+10 case (dots with error bars), compared to the results from \citet{hut93} (connected lines) over the range in common with our parameters. The cross section for exchange is shown in black, and the cross section for a resonant interaction is shown in red.  Error bars from the Hut paper are not shown, but they are slightly smaller than ours. The gray dashed line included for reference and is the same one shown in the Hut paper, which is analytic expression for the high-velocity limit of the ionisation cross section, given by  $\sigma = (40/9) \pi a^2/v^{-2}$ \citep{hut83b}.}
 \label{crosssection}
\end{figure}

Over most of our velocity range, we sample out to large enough $r_p$ that no more exchanges or resonances are occurring, i.e. the encounters are all flybys. In this case we can regard our experiments as fully sampling the parameter space of interest. At the lowest velocities we may be missing some very large $b$ encounters that could contribute to the cross sections, but the Hut data ends before that point anyway and no comparison to previous work can be made below $v_{\infty} / v_c \sim 0.08$. The comparison is shown in figure \ref{crosssection}, with the cross sections normalized to the area of the binary orbit with semi-major axis $a$. Although our error bars are slightly larger than the earlier work, which used about twice times as many encounters optimized to this calculation, the results of our experiments agree at about the $1\sigma$ level.

\bsp

\label{lastpage}

\end{document}